\documentclass{PoS}
\usepackage{amsmath,graphicx}
\title{Static-light hadrons on a dynamical anisotropic lattice}
\ShortTitle{Static-light hadrons on a dynamical anisotropic lattice}
\author{\speaker{Justin Foley}, Alan \'O Cais, Mike Peardon, Sin\'ead M.~Ryan, Jon-Ivar Skullerud\\
        School of Mathematics, Trinity College, Dublin, Ireland\\
        Email:\email{fly@maths.tcd.ie}
       }

\abstract{We present preliminary results for the static-light meson and baryon
spectra for $N_f=2$ QCD. The study is performed on an anisotropic lattice and 
uses a new all-to-all propagator method allowing us to determine particle masses to a 
high precision.}

\FullConference{XXIIIrd International Symposium on Lattice Field Theory\\25-30 July 2005\\Trinity College, Dublin, Ireland}

\PoS{PoS(Lat2005)216}

\begin{document}

\section{Introduction}
The simulation of hadrons containing one or more heavy quarks has been at the forefront of lattice QCD calculations 
for many years. Such hadrons can be problematic for lattice calculations because systematic errors which scale with the 
quark mass can make simulations with isotropic relativistic actions unreliable. The static limit, NRQCD and the 
Fermilab approach have all been used for simulations with $b$ quarks as have simulations at lighter quarks which 
are then extrapolated to the bottom quark mass. Results from the static limit can also be combined with those of 
lighter quarks to make an interpolation to the $b$ quark mass. 

The static limit, where $m_Q\rightarrow\infty$, is the lowest-order term in a $1/m_Q$ expansion of the 
Heavy Quark Effective Theory (HQET) Lagrangian. In this limit the approximate heavy quark spin and flavour 
symmetries become exact. It is believed that calculations in this limit can be relevant to $b$-physics; however, 
historically the signal-to-noise ratio in static-light correlation functions has been particularly poor. All-to-all 
propagators have been used in previous static-light simulations since they lead to an increase in statistics by placing 
the source and sink operators at every spatial site on the lattice~\cite{Michael:1998sg,Green:2003zz}. 
More recently, in 
Ref~\cite{DellaMorte:2003mn} a modified static quark action has been shown to give improved 
discretisation errors and signal-to-noise ratios in correlation functions. 

In this work, all-to-all propagators are used on anisotropic lattices for optimal determination of the particle 
energies. This algorithm has been described in detail in Ref~\cite{Foley:2005ac}. 

A 3+1 anisotropic action is used with $N_f=2$. The gauge action is a two-plaquette Symanzik-improved 
action which is described in Ref~\cite{Morningstar:1999dh}. The quark action has been designed 
for large anisotropies. The usual Wilson term appears in the fine temporal direction and doublers are 
removed in the coarse spatial directions by a Hamber-Wu term. Further details can be found in 
Ref~\cite{Foley:2004jf}. This action is written
\begin{equation}
S_q = \bar{\psi}\left(\gamma_0\nabla_0+\sum_i\mu_r\gamma_i\nabla_i\left( 1-\frac{1}{6}a_s^2\Delta_i\right) 
      -\frac{ra_t}{2}\Delta_0+sa_s^3\sum_i\Delta_i^2+m_0\right)\psi .
\end{equation}
where $\mu_r=(1+\frac{1}{2}ra_tm_0)$, $r=1$ is the usual coefficient of the Wilson term in the temporal direction 
and $s=\frac{1}{8}$ is the analogous coefficient in the spatial directions.  
%%%
\section{The static-light spectrum}
The static-light spectrum of hadrons can be classified according to the angular momentum of the 
light degrees of freedom, $J_l$. For static-light mesons the $S$, $P$ and $D$ wave states have
\begin{equation}
J_l = \underbrace{\frac{1}{2}^-,}_{\mathbf S}\; \underbrace{\frac{1}{2}^+,\; 
      \frac{3}{2}^+}_{\mathbf P},\; \underbrace{\frac{3}{2}^-,\; \frac{5}{2}^-}_{\mathbf D} .
\end{equation}
Interpolating operators are constructed for the light degrees of freedom which transform 
according to the irreducible representations of $O_h$. The relation between these irreducible 
representations and continuum quantum numbers can be determined by subduction. Table~\ref{tab:IrrepsofOh} lists 
the irreducible representations of the proper subgroup $O$ as well as the angular momenta of the lower-lying corresponding 
continuum states. 
\begin{table}[h]
\begin{center}
\begin{tabular}{c|c|c}
\hline
Lattice Irrep & Dimension & Continuum Irreps \\
\hline
\hline
$A_1$ & 1 & $0, 4, \ldots$ \\
$A_2$ & 1 & $3, \ldots$ \\
$E$   & 2 & $2, 4, \ldots$ \\
$T_1$ & 3 & $1, 3, \ldots$ \\
$T_2$ & 3 & $2, 3, \ldots$ \\
\hline
$G_1$ & 2 & $\frac{1}{2}, \frac{7}{2}, \ldots$ \\
$G_2$ & 2 & $\frac{5}{2}, \frac{7}{2}, \ldots$ \\
$H$   & 4 & $\frac{3}{2}, \frac{5}{2}, \ldots$ \\
\hline
\end{tabular}
\caption{The irreducible representations of the lattice rotation group, $O$ and their continuum analogues.}
\label{tab:IrrepsofOh}
\end{center}
\end{table}
Note that, for example, to study the $D$-wave mesons requires operators in the $G_2$ and $H$ irreducible 
representations. To construct the operators needed in a 
study of $S, P$ and $D$ wave states single-link and planar-diagonal displacements were used. 
\section{Results}
The preliminary results described here were obtained on dynamical lattices with
 $N_f=2$ and stout-link smeared gauge configurations~\cite{Morningstar:2003gk}. 
All-to-all propagators~\cite{Foley:2005ac,O'Cais:2004ww} with two sets of time-diluted noise 
vectors were used to improve the resolution of signals and enable us to determine 
both mesons and baryons. This lowest level of dilution was sufficient to 
obtain good signals in our static-light simulation. Five levels of Jacobi smearing were applied 
to the light quark fields and the two-point static-light correlation function was subsequently optimised 
using variational techniques. 

\begin{figure}[h]
\centering
\includegraphics*[width=6.5cm]{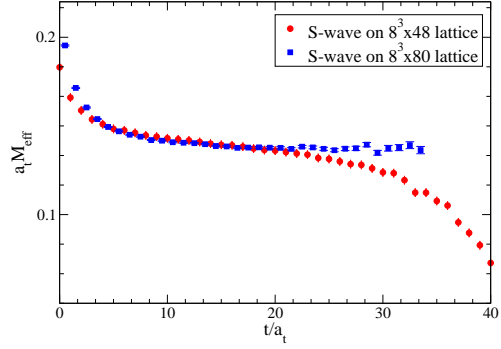}
\caption{A comparison of the effective mass of a static-light meson on $8^3\times 80$ and $8^3\times 48$ lattices. 
The longer lattice allows a more reliable determination of the ground state. }
\label{fig:short-lattice}
\end{figure}

The simulation parameters for this study are given in Table~\ref{tab:params}. 
The ratio of scales $\xi = a_s/a_t$ is 6 and the intricacies of non-perturbatively tuning the 
gauge and fermion anisotropy, $\xi$, are discussed in Ref~\cite{Morrin:2005tc}.
%%%
\begin{table}[h]
\begin{center}
\begin{tabular}{c|c}
\hline
\hline
Volume & $8^3\times 80$ \\
Configurations & 176 (mesons); 136 (baryons) \\
$\beta$ & 1.5 \\
$a_s$ & 0.21fm \\
$a_tm_{\rm sea}, a_tm_{\rm light} $ & -0.057 \\
$m_\pi /m_\rho$ & 0.55\\
\hline
\end{tabular}
\caption{Simulation parameters. The light and sea quark masses are close to the strange quark and their negative value in 
this table is an artefact of the Wilson additive quark mass renormalisation.}
\label{tab:params}
\end{center}
\end{table}
The results presented here are determined on a lattice extended in the temporal direction. This is motivated by 
our initial studies of the static-light mesons on an $8^3\times 48$ lattice. The quality of the data is now so good 
on these lattices that a very slowly falling signal was observed. This could not have been resolved with larger 
statistical errors. The comparison of data from the $8^3\times 48$ and the $8^3\times 80$ lattice is presented in 
Figure~\ref{fig:short-lattice}. The plot clearly shows that the longer lattice allows for a more reliable determination 
of the energy. The statistical errors are less than a percent in the region where a plateau is observed. 
\subsection{Mesons}
\begin{figure}[h]
\centering
\includegraphics*[width=6.5cm]{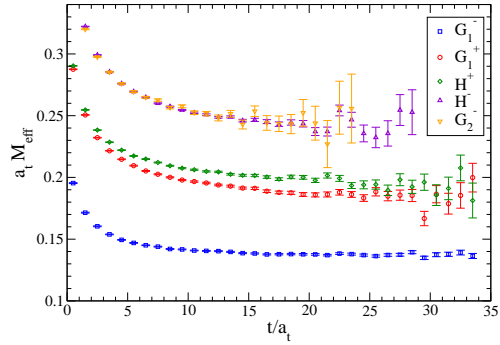}
\caption{Effective masses of the orbitally-excited static-light meson. States are labelled as in 
         Table 1 and by the parity of the meson.}
\label{fig:mesons}
\end{figure}

\begin{figure}
\centering
\includegraphics*[width=6.5cm]{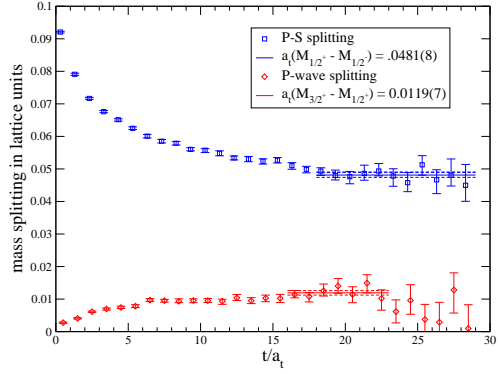}
\caption{The mass-splitting between $S$ and $P$ wave states (blue points) and between the $\frac{3}{2}^+$ and 
$\frac{1}{2}^+$ P wave states (red points). The solid lines are best fits to the splittings (not the difference of fits to 
effective masses).}
\label{fig:splittings1}
\end{figure}

\begin{figure}
\centering
\includegraphics*[width=6.5cm]{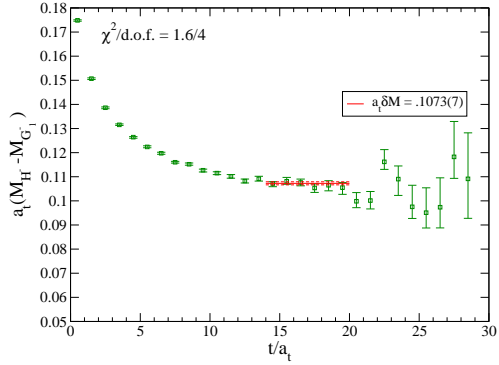}
\caption{The mass-splitting between the $D$ and $S$ waves, labeled $H^-$ and $G_1^-$ respectively.}
\label{fig:splittings2}
\end{figure}

Figure~\ref{fig:mesons} shows the effective masses of the static-light meson $S$, $P$ and $D$ waves. The states 
are labelled by their lattice irreducible representations and the corresponding continuum spin-states are listed in 
Table~\ref{tab:IrrepsofOh}. The plot demonstrates that all-to-all propagators yield a dramatic improvement 
in signal quality and allow the resolution of excited states. 
In the static limit mass splittings can be straightforwardly compared to their continuum values. In this study 
we find that these splittings can be determined very precisely. Figure~\ref{fig:splittings1} shows our preliminary
results for the $P-S$ splitting and the splitting between the $\frac{3}{2}^+$ and $\frac{1}{2}^+$ P-wave states.
We find that the $\Delta M(P-S)=0.0481(8)$ and 
$\Delta M(P_{\frac{3}{2}^+} - P_{\frac{1}{2}^+}) = 0.0119(7)$.
Figure~\ref{fig:splittings2} shows preliminary results for the $D-S$ splitting. 
The states are labelled by the irreducible representations of $O$ (as in Table~\ref{tab:IrrepsofOh})
and the parity of the meson. Within errors, no splitting is observed between the lowest-lying states 
in the $G_2^-$ and $H^-$ channels. This may indicate the inversion of the natural ordering of the $D$-wave multiplet predicted 
by some quark models, implying that the $\frac{3}{2}^-$ D-wave is in fact heavier than the $\frac{5}{2}^-$ D-wave. 
This has not been observed in other lattice studies. 
However, these are preliminary results and a conclusive statement can only be drawn when we 
have analysed the systematic errors including finite volume and discretisation effects. This is work in progress. 
The minimum timeslices used in Figures~\ref{fig:splittings1} and~\ref{fig:splittings2} is chosen by demanding that the 
effective masses of the particles involved have also reached a plateau at this time. 
\subsection{Baryons}
\begin{figure}[h]
\centering
\includegraphics*[width=6.5cm]{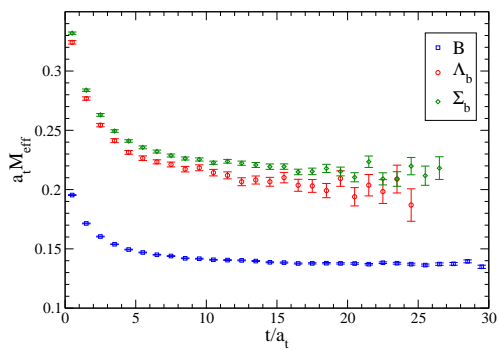}
\caption{The effective masses of the isoscalar ($\Lambda_b$) $\frac{1}{2}^+$ and 
         the isovector ($\Sigma_b$) $\frac{3}{2}^+$ baryons.}
\label{fig:baryons}
\end{figure}

Using all-to-all propagators we have constructed local operators for the isoscalar ($\Lambda_b$) $\frac{1}{2}^+$ and 
the isovector ($\Sigma_b$) $\frac{3}{2}^+$ baryons. The projectors, $\frac{1}{2}(1+\gamma_0)$ and 
$\frac{1}{2}(1-\gamma_0)$, applied to the light quark fields, as well as time-reversal symmetry, were used to improve signals 
in fits to the effective masses. Figure~\ref{fig:baryons} shows the effective masses of the $\Lambda_b$ and $\Sigma_b$. 
The effective mass of the $B$ meson is included for reference. The mass splittings are $a_t\Delta = 0.68(2)$ and 
$a_t\Delta = 0.078(14)$ for the $\Lambda_b$ and $\Sigma_b$ resectively. 
\section{Conclusions and outlook}
In this paper we have presented preliminary results for the masses and mass-splittings of static-light mesons and baryons, 
with $N_f=2$.
Using all-to-all propagators and anisotropic lattices, excellent signals for 
the $B$-meson and its radial excitations: $P$ and $D$ waves are found. 
Mesons are determined from 176 configurations while 136 configurations are
 used to study baryons. Even for this preliminary dataset, splitting between 
$P$ and $S$ waves, between $P$ waves and between $D$ and $S$ waves can be 
determined at the percent level. Our results also suggest an inversion in 
the natural ordering in the $D$-wave multiplet. This warrants further study and 
is work in progress. 
The masses of the $\Lambda_b$ and $\Sigma_b$ baryons and their splittings are 
also determined.
These results are very encouraging and we plan to extend this study to larger 
volumes with correctly renormalised parameters, $\xi_q$ and $\xi_g$. 
\section*{Acknowledgments}
This work is supported by IRCSET Embark Initiative awards SC/03/393Y and RS/2002/208-7M, 
and the IITAC PRTLI initiative.

\bibliography{proceedings}

\end{document}